\documentclass[aip,jcp,reprint]{revtex4-1}
\usepackage{amsmath}
\usepackage{esint}
\usepackage{color}

\begin{document}

\title{An alternative, dynamic density functional-like theory for time-dependent
density fluctuations in glass-forming fluids} 



\author{Grzegorz Szamel}
\email[Email: ]{grzegorz.szamel@colostate.edu}
\affiliation{Department of Chemistry, Colorado State University, Fort Collins, 
Colorado 80523, USA}

\date{\today}

\begin{abstract}
We propose an alternative theory for the relaxation of density fluctuations in 
glass-forming fluids. We derive an equation of motion for the density 
correlation function which is local in time and is similar in spirit to the equation 
of motion for the average non-uniform density profile derived within  
the dynamic density functional theory. We identify the Franz-Parisi free
energy functional as the non-equilibrium free energy for the evolution of the
density correlation function. An appearance of a local minimum of this functional 
leads to a dynamic arrest. Thus, the ergodicity breaking transition predicted by 
our theory coincides with the dynamic transition of the static approach based on 
the same non-equilibrium free energy functional.
\end{abstract}

\maketitle 

\section{Introduction}\label{intro}

Slow dynamics in glass-forming systems is usually monitored through time-dependent
density correlation functions \cite{BerthierBiroliRMP}. 
The commonly used theoretical description of the density 
correlation function starts with the memory function representation, which absorbs
the complicated dynamics of density fluctuations into a time-delayed
friction kernel called memory function \cite{Zwanzigbook}.
The exact expression for the memory function involves a pair-density correlation
function evolving with the so-called projected dynamics \cite{Goetzebook} which
is difficult to analyze or even simulate \cite{Marco}. 
To proceed, a factorization approximation
is used which expresses the pair-density correlation function evolving with projected
dynamics as a product of two density correlation functions evolving with 
standard dynamics \cite{Goetzebook}. 
The resulting closed set of equations constitutes the 
mode-coupling theory of glass-forming liquids
\cite{Goetzebook,DasRMP,RC}, which
describes well several features of glassy 
dynamics. Its most often quoted success is the 
description of the cage effect, which manifests itself through a 
characteristic two-step decay of the density correlation function, 
with an intermediate-time plateau whose extension grows upon cooling. The theory 
reproduces well the wavevector dependence of the plateau  and 
non-trivial power laws describing the approach to and the departure 
from the plateau \cite{Goetzebook}. The factorization
approximation can be pushed to higher-level correlation functions, leading to one of 
generalized mode-coupling theories \cite{Szamel2003,WuCao,Janssen2015}. Qualitatively, 
predictions of these theories are similar to those of the standard mode-coupling
theory. However, these more sophisticated theories significantly improve
upon the standard theory: they describe well the time-dependence of the
density correlation function for a larger range of control 
parameters \cite{Janssen2015}. 

One common feature of all mode-coupling-like approaches is the prediction of an 
ergodicity breaking transition, referred to as the mode-coupling transition, 
at which the plateau of the density correlation function extends infinitely. 
In two- or three-dimensional systems the mode-coupling 
transition is 
replaced by a smooth crossover. 
The dynamics beyond the crossover is called 
``activated'' \cite{BerthierBiroliRMP} but there is no commonly 
accepted quantitative description of the postulated activated processes 
\cite{Note1}. 

In addition to the fundamental problems of the avoided transition and the activated
dynamics, the standard mode-coupling theory and its generalizations 
suffer from another important conceptual problem. Some time after the mode-coupling
theory was formulated a number of static descriptions of the glass formation
were proposed \cite{MP1,MP2,PZ1,PZ2}. 
Most of these descriptions rely upon a free-energy-like functional
derived using the replica method. They predict a transition which,
on the basis of the analogy with the results obtained for exactly solvable spin-glass 
models, is referred to as the dynamic transition. At this transition non-trivial
replica off-diagonal density correlations appear which are conceptually identified 
with the plateau of the time-dependent correlation function at the mode-coupling 
transition \cite{PZ2,JacquinZamponi}. 
However, the location of the dynamic transition and the wavevector 
dependence of the replica off-diagonal density correlations are different from
the location of the mode-coupling transition and the wavevector 
dependence of the infinite time plateau, respectively. 
In particular, in static theories of the glass formation 
replica off-diagonal correlations are obtained by minimizing a free-energy-like 
functional. In contrast, the self-consistent equation 
for the wavevector dependence of the plateau derived within mode-coupling approaches 
cannot be obtained by a functional minimization. Thus, the two approaches are 
fundamentally incompatible\cite{Note2}. This is a major conceptual problem since the 
static and dynamic (mode-coupling) approaches are considered to be the two 
facets of a unified description of glassy phenomena referred to as random 
first-order transition theory \cite{RFOTorg,RFOTrec}. 

Recall that in the limit of large spatial dimensions a 
consistent description of the glass formation and glassy dynamics has been 
formulated \cite{PUZbook}. 
Both exact static \cite{highdstatic} and dynamic \cite{highddynamic} 
theories have been developed and
it was shown that the dynamic transition predicted by the static approach coincides
with the 
mode-coupling-like transition predicted by the dynamic
theory \cite{PUZbook}. 

Importantly, while there exist static approaches that with increasing spatial 
dimension reproduce the exact large dimensional static theory, in the same limit the 
mode-coupling theory becomes drastically different from the exact 
large dimensional 
dynamic theory \cite{SchmidSchillingPRE,IkedaMiyazakiPRL2010}. 
To rationalize this fact we recall 
that the mode-coupling theory and the exact large dimensional 
dynamic theory, while similar in spirit, are nevertheless technically quite different. 
In particular, within the mode-coupling theory one derives a self-consistent
equation for the time-dependent density correlation function whereas the
exact large dimensional dynamic theory imposes a self-consistency condition
on a stochastic process \cite{Manacorda}. This condition cannot be expressed in terms of
a self-consistent equation for a correlation function.

The above described unsatisfactory state of the finite-dimensional dynamic theory
calls for a reformulation of the mode-coupling approach and/or for exploration of
other approaches to the dynamics of glassy fluids. In this contribution we propose
a possible alternative approach. Our theory is
consistent with static approaches that rely upon free-energy-like functionals. 
We identify a functional generalization of the Franz-Parisi potential \cite{FP1,FP2},
which we call the Franz-Parisi functional, as the free-energy-like object 
generating the time dependence of the density correlation function. The equation
of motion that we derive is similar in spirit to the equation of motion
derived within dynamic density functional theory \cite{DDFTreview}. 
It is local in time and thus it does not capture 
the time-delayed friction which is considered the 
central feature of glassy dynamics. We believe that the present equation is
just the first ingredient of a new dynamic theory and that the time-delayed 
friction can be recovered if an analogue of the memory function is identified.

In the following we present a somewhat heuristic derivation of our theory.
A more formal, projection operator-based, derivation and incorporation of
time-delayed friction are left for future work.

\section{Formulation of the problem}

We consider a system of $N$ particles in volume $V$, with pair-wise
additive interactions determined through spherically-symmetric potential $V(r)$.
We assume that the microscopic dynamics is Brownian \cite{Note3}. We denote the 
diffusion and friction coefficients of an isolated particle by $D_0$ and
$\gamma$, respectively, with  $D_0 = T/\gamma$, where $T$ is the temperature
and the Boltzmann constant is set to 1.

We aim to develop a theory for the density correlation function in the 
Fourier space, $F(k;t)$,
\begin{eqnarray}\label{dcf1}
F(k;t) = N^{-1} \left<n(\mathbf{k})e^{\Omega t} n(-\mathbf{k}) \right>.
\end{eqnarray}
Here $n(\mathbf{k}) = \sum_j e^{-i\mathbf{k}\cdot\mathbf{r}_j}$ is the Fourier
transform of the microscopic density and $\Omega$ is the $N$-particle 
evolution operator, \textit{i.e.} the Smoluchowski operator,
\begin{eqnarray}\label{Omegadef}
\Omega = -D_0  \sum_j \partial_{\mathbf{r}_j}\cdot \left[
- \partial_{\mathbf{r}_j} + \beta \mathbf{F}_j
\right],
\end{eqnarray}
where $\beta=1/T$ and $\mathbf{F}_j$ is the force on particle $j$,
$\mathbf{F}_j= \sum_{l\neq j} \mathbf{F}(\mathbf{r}_{jl})$ with  
$\mathbf{F}(\mathbf{r}_{jl})= - \partial_{\mathbf{r}_j} V(r_{jl})$.
Finally, $\left<\ldots\right>$ in Eq. \eqref{dcf1} denotes the equilibrium average;
we use the convention that the probability distribution stands to the right of the 
quantity being averaged, and all operators to its left act on it as well as on
everything else.

The standard approach to develop a theory for the density correlation
function starts with the projection operator manipulations \cite{SL}. Below we present 
an equivalent formulation of the first step of the standard approach, which avoids an 
explicit introduction of a projector operator. This formulation is inspired by 
the so-called linear kinetic theory developed to describe time-dependent equilibrium
correlation functions \cite{ResiboisLebowitz,BlawzdzCichocki,SzamelLeegwater}. 
We will contrast this approach with our new theory in Sec. \ref{qeapp}. 

We start by recognizing that the right-hand-side of Eq.  
\eqref{dcf1} can be interpreted as the density of a $N$-particle system that
has time-dependent ``probability distribution'' of the following form,
\begin{equation}\label{Plinex}
P_\text{l}(t) = e^{\Omega t} n(-\mathbf{k}) P_\text{eq},
\end{equation} 
where $P_\text{eq}$ is the equilibrium probability distribution. We used quotation
marks to emphasize that strictly speaking $P_\text{l}(t)$ is not a probability 
distribution; in particular, it is not properly normalized. However, as explained
by R\'{e}sibois and Lebowitz \cite{ResiboisLebowitz}, 
``formally, this makes no difference''. Subscript
``l'' in $P_\text{l}(t)$ indicates that distribution \eqref{Plinex} has the same form as 
the \emph{linear} change of the $N$-particle distribution due to an external potential
that is turned off at the initial time, $t=0$.   

Next, we write the equation of motion for the density correlation function
\begin{eqnarray}\label{dcfeom1}
\partial_t F(k;t) = N^{-1} \left<n(\mathbf{k}) \Omega \right>_{\text{l}},
\end{eqnarray}
where $\left<\ldots\right>_{\text{l}}$ indicates averaging with time-dependent 
distribution $P_\text{l}(t)$, Eq. \eqref{Plinex}. 
We recall that following our convention, the evolution 
operator acts on distribution \eqref{Plinex}, which stands to the right 
of $\Omega$.

To calculate the right-hand-side of Eq. \eqref{dcfeom1} we assume that
at later times, $t>0$, distribution $P_\text{l}(t)$ has the same form as at the
initial time but with a different magnitude,
\begin{equation}\label{Plinapp1}
P_\text{l}(t)\approx P_\text{l}^\text{a}(t)\equiv n(-\mathbf{k}) f(t) P_\text{eq}
\end{equation} 
where superscript ``a'' indicates the approximate character of distribution 
$P_\text{l}^\text{a}$ 
and function $f(t)$ is chosen in such a way that averaging with $P_\text{l}^\text{a}(t)$ 
reproduces the density correlation function at time $t$,
\begin{equation}\label{Plinapp2}
N^{-1} \left<n(\mathbf{k})\right>_{\text{l}}^\text{a} = F(k;t).
\end{equation}
In Eq. \eqref{Plinapp2} 
$\left<\ldots\right>_\text{l}^\text{a}$ indicates averaging with 
$P_\text{l}^\text{a}(t)$.

A simple calculation gives 
$f(t) = F(k;t)/S(k)$, where $S(k)$ is the static structure factor,
$S(k)= N^{-1} \left<n(\mathbf{k}) n(-\mathbf{k}) \right>$,
and approximate distribution \eqref{Plinapp1} which
depends \textit{linearly} on the density correlation function
at time $t$, $P_\text{l}^\text{a}(t) 
= n(-\mathbf{k}) \left(F(k;t)/S(k)\right) P_\text{eq}$.

Using distribution $P_\text{l}^\text{a}$
in Eq. \eqref{dcfeom1} we obtain 
the following approximate equation of motion for the density correlation function,
\begin{eqnarray}\label{dcfeom2}
\partial_t F(k;t) = - D_0 \left(k^2/S(k)\right) F(k;t). 
\end{eqnarray}
The same equation is obtained using the first step of the standard projection
operator procedure, \textit{i.e.} when one neglects the memory function term. 
We note that equation of motion \eqref{dcfeom2} is local in time
and predicts that different Fourier components of $F(k;t)$ relax independently.
Only after the memory function is included, the relaxation of different Fourier 
components (modes) becomes coupled. This fact was the motivation for the name
``mode-coupling'' given to the first theory of this type, which was developed 
to describe critical dynamics \cite{Kawasaki}. 

\section{An alternative formulation of the theory for density correlation function}

We propose the following alternative approach to approximately evaluate the time
evolution of the density correlation function. First, we re-write definition
\eqref{dcf1} of the density correlation function by distinguishing between
averaging over the time evolution (which corresponds to averaging over noise
in the Langevin formulation of Brownian dynamics) and averaging over
the initial positions of the particles that will be denoted by $\mathbf{r}_j^0$, 
$j=1,\ldots,N$, 
\begin{eqnarray}\label{dcf2}
F(k;t) = N^{-1} \left<\left<n(\mathbf{k}) \right>_\text{n} 
n^0(-\mathbf{k})\right>_0.
\end{eqnarray}
Here $\left<\ldots\right>_\text{n}$ denotes averaging with the following 
time-dependent distribution 
\begin{equation}\label{Pnlex}
P_\text{n}(t) = e^{\Omega t} \left(N!\right)^{-1} 
\sum_\pi \prod_{j} \delta(\mathbf{r}_j-\mathbf{r}_{\pi(j)}^0),
\end{equation}
where $\sum_\pi$ denotes sum over permutations of particle labels,
$\pi(j)$ is the label of particle $j$ in permutation $\pi$ and 
subscript ``n'' emphasizes that distribution $P_\text{n}(t)$ \emph{cannot}
be interpreted as a linear change from the equilibrium distribution. 
We note that $P_\text{n}(t)$ is the distribution of particles' positions 
at time $t$, which are denoted by $\mathbf{r}_j$, $j=1,\ldots,N$, but it
depends parametrically on the initial positions $\mathbf{r}_j^0$, 
$j=1,\ldots,N$. Furthermore,
in Eq. \eqref{dcf2} $n^0(-\mathbf{k})$ 
denotes the microscopic density calculated for the initial configuration, 
$n^0(\mathbf{k}) = \sum_j e^{-i\mathbf{k}\cdot\mathbf{r}_j^0}$
and $\left<\ldots\right>_0$ denotes averaging over the equilibrium distribution
of the initial positions of the particles. 

Eqs. (\ref{dcf2}-\ref{Pnlex}) allow for the following procedure to evaluate the
density correlation function. First, one calculates average density, 
$\left<n(\mathbf{k}) \right>_\text{n}$, using distribution \eqref{Pnlex}. This 
average density depends on the initial positions of all the particles. Then, 
one calculates the joint average of $\left<n(\mathbf{k}) \right>_\text{n}$ 
and density configuration $n^0(-\mathbf{k})$ over the equilibrium distribution of
the initial positions. 

The advantage of this formulation is that to describe the evolution of density 
$\left<n(\mathbf{k}) \right>_\text{n}$ one can (in fact, one should) go beyond 
the expansion used in the linear kinetic theory. In other words, the approximate
theory for $\left<n(\mathbf{k}) \right>_\text{n}$ should be non-linear. 

Since one has to calculate
the time evolution of the average density, one could try using 
the dynamic density functional theory \cite{DDFTreview}. However,
one cannot use the standard version of this approach 
\cite{MarconiTarazona,ArcherEvans,EspanolLoewen} 
since the important feature of average density 
$\left<n(\mathbf{k}) \right>_\text{n}$ is that it depends on 
the initial positions of all the particles, \textit{i.e.} on $N$ parameters, 
$\mathbf{r}_j^0$, $j=1,\ldots,N$. 

Our proposed approach is similar in spirit to
that used by Dufty and Rodriguez \cite{DuftyRodriguez} to 
elucidate an earlier 
result due to Hauge \cite{Hauge}. Hauge
showed that one can obtain long-time mode-coupling contributions
to time-dependent equilibrium correlation functions from the non-linear 
Boltzmann equation. This was un-expected since mode-coupling contributions 
were thought to originate from correlated sequences of particles' interactions 
whereas the Boltzmann equation was known to include only uncorrelated collisions.
Dufty and Rodriguez pointed out that Hauge's result could be explained if 
his one-particle density 
were re-interpreted 
as a 
density that implicitly depends on the positions of all the particles. 
They showed that for the hard-sphere 
system this new density satisfies exactly a non-linear equation that has the same 
form as the Boltzmann equation. Once the time dependence of the new density 
is evaluated, its correlation
with the density of the initial configuration reproduces the exact time-dependent
equilibrium correlation function.  
Our average density $\left<n(\mathbf{k}) \right>_\text{n}$
is analogous to the one-particle density introduced by Dufty and Rodriguez. 

\section{Quasi-equilibrium approximation}\label{qeapp}

To proceed, we write down equation of motion 
for the density correlation function
\begin{eqnarray}\label{dcfeom3}
\partial_t F(k;t) = \left<\left<n(\mathbf{k}) \Omega  \right>_\text{n}
n^0(-\mathbf{k})\right>_0.
\end{eqnarray}
To evaluate the right-hand-side of Eq. \eqref{dcfeom3} we need an approximate
expression for $P_\text{n}(t)$. 

At this point we recall the 
central assumption of the dynamic density functional theory, which is known as the
\emph{adiabatic approximation} \cite{DDFTreview,ArcherEvans}. 
It states that the correlations in a non-equilibrium system are 
the same as those in an equilibrium system with a non-uniform 
density equal to the instantaneous density of the non-equilibrium system. 
We follow this approximation in spirit and assume that non-equilibrium
correlations embodied in distribution \eqref{Pnlex} are the same as in
an equilibrium system in which averaged density averaged over the initial
conditions is uniform and the correlation of the averaged density with the density
of the initial configuration reproduces the time-dependent density correlation
function. 
 
Specifically, we propose the following approximate formula for $P_\text{n}(t)$,
\begin{widetext}
\begin{eqnarray}\label{Pnlapp1}
P_\text{n}(t) 
\approx  P_\text{n}^\text{a}(t) \equiv 
\frac{\exp\left[-\beta \sum_{j\neq l} V(r_{jl})\right]}
{Z[V^\text{ext}_1,V^\text{ext}_2]} 
\exp\left[-\beta\sum_j V_1^\text{ext}(\mathbf{r}_j) 
-\beta\sum_{j,l} V_2^\text{ext}(|\mathbf{r}_j-\mathbf{r}^0_l|)\right],
\end{eqnarray}
where $Z[V^\text{ext}_1,V^\text{ext}_2]$ is the partition function,
\begin{eqnarray}\label{Z}
Z[V^\text{ext}_1,V^\text{ext}_2] = \int d\mathbf{r}_1 \ldots d\mathbf{r}_N 
\exp\left[-\beta \sum_{j\neq l} V(r_{jl})-\beta\sum_j V_1^\text{ext}(\mathbf{r}_j) 
-\beta\sum_{j,l} V_2^\text{ext}(|\mathbf{r}_j-\mathbf{r}^0_l|)\right],
\end{eqnarray}
\end{widetext}
and where one- and two-body time-dependent 
external potentials, $V^\text{ext}_1$ and $V^\text{ext}_2$,
are determined by two conditions described above. Explicitly, we require that 
density $\left<n(\mathbf{k})\right>_\text{n}^\text{a}$, 
where $\left<\ldots\right>_\text{n}^\text{a}$ denotes averaging with distribution
\eqref{Pnlapp1}, is on average uniform,
\begin{equation}\label{cond1}
N^{-1}\left<\left<n(\mathbf{k})\right>_\text{n}^\text{a}\right>_0 = 
\delta_{\mathbf{k},0}.
\end{equation}
Second, we require that the correlation of density
$\left<n(\mathbf{k})\right>_\text{n}^\text{a}$ 
and the density of the initial configuration
reproduces the density correlation function at time $t$,
\begin{equation}\label{cond2}
N^{-1}\left<\left<n(\mathbf{k})\right>_\text{n}^\text{a}n^0(-\mathbf{k})\right>_0 
= F(k;t).
\end{equation}
We assume that conditions (\ref{cond1}-\ref{cond2}) uniquely determine
potentials $V^\text{ext}_1$ and $V^\text{ext}_2$. 

Physically, two-body potential $V^\text{ext}_2$ is the 
interaction necessary to maintain the correlation between the state of the system
at time $t$ and the initial configuration that is equal to $F(k;t)$, while
keeping the uniform average density, which is maintained by the 
additional one-body potential. We emphasize that approximate instantaneous distribution
\eqref{Pnlapp1} depends on the density correlation function at time $t$
in a \textit{non-linear} and complicated way.

We note that in the first step of the two-step averaging process, \textit{i.e.} 
while evaluating 
$\left< \ldots \right>_\text{n}^\text{a}$, 
the initial positions $\mathbf{r}^0_j$, $j=1,\ldots,N$, play the role of 
the quenched variables. Within our theory they appear in a very natural way.

We propose the name \emph{quasi-equilibrium approximation} for formula \eqref{Pnlapp1} to 
emphasize that the distribution of particle positions at time $t$ is the same as 
in an equilibrium state that satisfies conditions (\ref{cond1}-\ref{cond2}). 
The additional motivation for this name is the conceptual similarity of our 
approximation with 
quasi-equilibrium construction for the long-time 
glassy dynamics proposed by Franz \textit{et al.} \cite{FranzQE1,FranzQE2}. We 
comment on this point in the Discussion.

The final step is the evaluation of the average at the right-hand-side of 
Eq. \eqref{dcfeom3} with the approximate distribution \eqref{Pnlapp1} in
terms of reduced distribution functions and external two-body potential 
$V^\text{ext}_2$. After some work we arrive at the following equation of 
motion
\begin{eqnarray}\label{dcfeom4}
&& \partial_t F(k;t) =
D_0 \int \frac{d\mathbf{k}_1}{(2\pi)^3} 
\mathbf{k}\cdot(\mathbf{k}-\mathbf{k}_1)
\\ \nonumber && \times
N^{-1} \left<\left< n(\mathbf{k}_1)
\right>_\text{n}^\text{a} n^0(-\mathbf{k})n^0(\mathbf{k}-\mathbf{k}_1)
\right>_0
\beta V^\text{ext}_2(|\mathbf{k}-\mathbf{k}_1|).
\end{eqnarray}
The right-hand-side of Eq. \eqref{dcfeom4} is written in terms of two objects,
a three-body average $\left<\left< n(\mathbf{k}_1)
\right>_\text{n}^\text{a} n^0(-\mathbf{k})n^0(\mathbf{k}-\mathbf{k}_1)
\right>_0$ and the Fourier transform of two-body potential $V^\text{ext}_2$. 
Both of these objects
are functionals of the instantaneous value of the density correlation function, 
$F(k;t)$. We discuss two possible technical approximations for 
the three-body average $\left<\left< n(\mathbf{k}_1)
\right>_\text{n}^\text{a} n^0(-\mathbf{k})n^0(\mathbf{k}-\mathbf{k}_1)
\right>_0$ in Appendix A. We show that for non-interacting particles
Eq. \eqref{dcfeom4} reproduces the known exact result in Appendix B.

According to equation \eqref{dcfeom4}, the driving force for the time evolution
of the density correlation function is the interaction necessary to maintain
instantaneous correlations between the density at time $t$ and the initial density.
We recall that at a dynamic transition of a static theory of the glass formation 
the correlation between the density of the system and the so-called template 
(also know as the zeroth replica) appears spontaneously, without any
system-template interaction. This implies that at the dynamic transition of a 
static theory the right-hand-side of Eq. \eqref{dcfeom4} vanishes and the
density correlation function freezes. 
In other words, Eq. \eqref{dcfeom4} predicts an ergodicity-breaking
transition that coincides with the dynamic transition of the static 
theory. 

To further develop the connection with static glass formation theories 
we note that partition function \eqref{Z} can be used to introduce a 
free-energy-like functional,
\begin{eqnarray}\label{F}
\mathcal{F}[V^\text{ext}_1,V^\text{ext}_2] = -T \ln Z[V^\text{ext}_1,V^\text{ext}_2].
\end{eqnarray}
At a given time, functional $\mathcal{F}$ depends on the initial positions of 
the particles, which play the role of quenched variables. It should be 
self-averaging with respect to the distribution of initial positions. 
By a Legendre transform of functional $\mathcal{F}$ with respect to the one- and
two-body potentials one can obtain a functional that depends on the average
density and average instantaneous value of the time-dependent density correlation
function. The latter functional 
is a functional generalization of the
Franz-Parisi potential \cite{FP1,FP2}. We believe that two-body potential 
$V^\text{ext}_2$ can be obtained as a functional derivative of the Franz-Parisi 
functional with respect to the instantaneous density correlation function, 
at constant average density. Thus, the time evolution predicted by 
Eq. \eqref{dcfeom4} stops when the Franz-Parisi functional reaches its local minimum. 

Finally, to make explicit contact with the dynamic density functional theory 
we recall that its equation for the time evolution of the non-uniform
average density $\bar{n}(\mathbf{k};t)$ can be written in a very similar 
way \cite{Note5},
\begin{eqnarray}\label{ddfteom}
\partial_t \bar{n}(\mathbf{k};t) =
D_0 \int \frac{d\mathbf{k}_1}{(2\pi)^3} 
\mathbf{k}\cdot(\mathbf{k}-\mathbf{k}_1) \bar{n}(\mathbf{k}_1;t) 
\beta V^\text{ext}(\mathbf{k}-\mathbf{k}_1).
\nonumber \\
\end{eqnarray}
In Eq. \eqref{ddfteom} $ V^\text{ext}$ is the external potential needed to
maintain non-uniform equilibrium density equal to the instantaneous 
average density $\bar{n}(\mathbf{k};t)$.  

\section{Discussion}

We proposed here an alternative theory for the relaxation
of density fluctuations in glassy fluids and the glass transition. 
The main approximation of our theory is that the correlations
between the state of the system at time $t$ and the initial state of the system
can be reproduced by a Boltzmann-like formula coupling these two systems in
a quasi-equilibrium fashion. The central quantity that allows one to calculate
the required coupling is a generalization of the Franz-Parisi potential 
which gives more freedom to the coupling between the system and the template. 
Our equation of motion for the density correlation function is local in time but
the relaxation of different Fourier components of the 
density correlation function is coupled. Our approach 
predicts an ergodicity-breaking transition identical to the dynamic transition 
predicted by a static theory of the glass formation based on the same 
Franz-Parisi functional. We believe that a numerical implementation
of our theory should start with a specific approximate Franz-Parisi functional
and use it to calculate the two-body potential needed to integrate Eq. 
\eqref{dcfeom4}. This task is is left for future research.

It would be interesting to test the present approach on 
a spherical $p$-spin model for which the time-dependence of correlation functions
can be analyzed exactly. 

We envision two directions to extend our theory. First, the present theory
uses an order parameter, density correlation function, that is uniform in space.
The time evolution of this order parameter stops at the ergodicity-breaking transition 
which corresponds to a local minimum of the Franz-Parisi functional. 
However, in finite dimensions
relaxation beyond the local minimum can happen via nucleation and growth processes.
To investigate such processes one needs to allow for inhomogeneous order parameter
fields, in the spirit of Franz \cite{Franz2005,Franz2007} and Wolynes \textit{et al.}
\cite{Dzero2005}. We also note that inhomogeneous order parameter fields were
observed in numerical investigation of Cammarota \textit{et al.} \cite{Cammarota2010}.
They were also introduced within dynamic field theory developed by 
Rizzo \cite{Rizzo2014,Rizzo2016}. 

Second, the local in time equation of motion implies that our theory misses time-delayed
friction. This is in contrast
to the mode-coupling approach, which prominently features non-local in time 
relaxation processes. We believe that a more formal derivation of our theory,
based on a Kawasaki-Gunton-style projection operator \cite{KawasakiGunton}, can
result in a generalization of equation of motion \eqref{dcfeom4}, which will 
include a memory function term describing time-delayed friction. 
Since the projection operator term will involve time-dependent
quasi-equilibrium distribution \eqref{Pnlapp1}, we anticipate that 
the memory function expression
will feature time-dependent vertices. We hope that as a result, the 
incorporation of the memory function
will modify the relaxation near the ergodicity-breaking transition but will
not change its location.  

In the context of the non-local in time relaxation processes we would like to comment 
on the relation of our approximation and the quasi-equilibrium construction
of Franz \textit{et al.} \cite{FranzQE1,FranzQE2} We assumed that to calculate 
the time derivative of the density correlation function we can approximate
the exact probability distribution \eqref{Pnlex} by an equilibrium distribution
in an external potential that assures that the correlation between the density 
of the system at time $t$ and its initial initial density is equal to $F(k;t)$.
In contrast, the starting assumption of Franz \textit{et al.} is a
quasi-equilibrium condition for the \textit{transition probability}. 
Franz \textit{et al.} showed that this assumption leads to non-local in time 
equations describing the long-time behavior of the density correlations in the 
vicinity of the plateau. 

Finally, we note that the present theory shares some features with the
co-called ``naive'' mode-coupling theory that is the starting point of the
non-linear Langevin equation theory of activated hopping proposed by Schweizer
and collaborators \cite{Schweizer2005,SchweizerSaltzman}. 
This theory's equation of motion for
the averaged order parameter is local in time, features a non-equilibrium 
free-energy-like function and predicts an ergodicity-breaking transition at the 
point at which this function develops a local minimum. The main difference
is that our order parameter is a collective quantity whereas that of 
Schweizer and collaborators \cite{Schweizer2005,SchweizerSaltzman} 
is a single-particle quantity. In addition, we anticipate 
that in our case barrier crossing is facilitated by non-uniform order
parameter fields whereas the theory of Schweizer
and collaborators \cite{Schweizer2005,SchweizerSaltzman}
implicitly assumes a uniform, thermally facilitated barrier hopping.

\vskip -2ex
\begin{acknowledgments}
I thank Elijah Flenner and Francesco Zamponi for comments on the manuscript. 
I gratefully acknowledge the support of NSF Grant No.~CHE 1800282.
\end{acknowledgments}

\vskip -2ex
\section*{Data availability}

Data sharing is not applicable to this article as no new data were
created or analyzed in this study.

\appendix
\numberwithin{equation}{section}
\section{Two possible approximations for
three-body average $\left<\left< n(\mathbf{k}_1)
\right>_\text{n}^\text{a} n^0(-\mathbf{k})n^0(\mathbf{k}-\mathbf{k}_1)
\right>_0$}

In a numerical implementation of our theory one would like to start with 
a scheme that uses only two-particle correlations. To this end
one would express  three-body average $\left<\left< n(\mathbf{k}_1)
\right>_\text{n}^\text{a} n^0(-\mathbf{k})n^0(\mathbf{k}-\mathbf{k}_1)
\right>_0$ in terms of the density correlation function and static pair
correlation functions. Here we propose two possible approximations,
which are linear in density correlation function. We note that the
right-hand-side on Eq. \eqref{dcfeom4} is still a non-linear functional
of $F(k;t)$ due to the presence of the two-body potential. 

First, one can argue that at short times the two-body potential
$V^\text{ext}_2$ is extremely short range and therefore in approximate expression
\eqref{Pnlapp1} a given $\mathbf{r}_j$ is correlated only with
one of $\mathbf{r}_l^0$s. This suggests that one may neglect off-diagonal terms
in $n^0(-\mathbf{k})n^0(\mathbf{k}-\mathbf{k}_1)$, which results in
\begin{eqnarray}\label{shorttime}
\left<\left< n(\mathbf{k}_1)
\right>_\text{n}^\text{a} n^0(-\mathbf{k})n^0(\mathbf{k}-\mathbf{k}_1)
\right>_0 \approx
\left<\left< n(\mathbf{k}_1)
\right>_\text{n}^\text{a} n^0(-\mathbf{k}_1)\right>_0. 
\nonumber \\
\end{eqnarray}

Alternatively, one may resort to a convolution-like approximation
\begin{eqnarray}\label{conv}
&& \left<\left< n(\mathbf{k}_1)
\right>_\text{n}^\text{a} n^0(-\mathbf{k})n^0(\mathbf{k}-\mathbf{k}_1)
\right>_0 \approx
\\ \nonumber && 
\left<\left< n(\mathbf{k}_1)\right>_\text{n}^\text{a} n^0(-\mathbf{k}_1)\right>_0 
S(k) S(|\mathbf{k}-\mathbf{k}_1|).
\end{eqnarray}
where we used the equilibrium 
distribution of the initial positions, which implies 
$N^{-1} \left<n^0(\mathbf{k}) n^0(-\mathbf{k})\right>_0 = S(k)$.

\section{Limiting case: non-interacting particles}

As a ``sanity check'', which was also performed while deriving dynamic density functional
theory \cite{MarconiTarazona}, we consider here Eq. \eqref{dcfeom4} 
for non-interacting particles. In this case the short-time approximation
\eqref{shorttime} is exact and Eq. \eqref{dcfeom4} in the direct space reads
\begin{eqnarray}\label{dcfeom5}
&& \partial_t N(|\mathbf{r}-\mathbf{r}^0|;t) = - D_0 
\partial_{\mathbf{r}} N(|\mathbf{r}-\mathbf{r}^0|;t) \partial_{\mathbf{r}}
\beta V^\text{ext}_2(|\mathbf{r}-\mathbf{r}^0|;t),
\nonumber \\
\end{eqnarray}
where $N(|\mathbf{r}-\mathbf{r}^0|;t)-n$ is the inverse Fourier transform
of $F(k;t)$. Then, one can show that for non-interacting particles
$\beta V^\text{ext}_2(|\mathbf{r}-\mathbf{r}^0|)=
-\ln\left[N(|\mathbf{r}-\mathbf{r}^0|;t)\right]$ and thus Eq. \eqref{dcfeom5}  
reproduces the exact equation of motion for density correlation function
of non-interacting particles. We note that the same equation of motion is obtained 
from Eq. \eqref{dcfeom2}, since for non-interacting particles $S(k)=1$.


\section*{References}

\begin{thebibliography}{99}
\bibitem{BerthierBiroliRMP} L. Berthier and G. Biroli,
Rev. Mod. Phys. \textbf{83}, 587 (2011).
\bibitem{Zwanzigbook} R. Zwanzig, \textit{Nonequilibrium Statistical
Mechanics}, (Oxford, New York, 2002).
\bibitem{Goetzebook}
W. G\"otze, \textit{Complex dynamics of glass-forming
liquids: A mode-coupling theory} (Oxford University Press, Oxford, 2008).
\bibitem{Marco} 
M. Baity-Jesi and D.R. Reichman, J. Chem. Phys. \textbf{151}, 084503 (2019).
\bibitem{DasRMP} S.P. Das, Rev. Mod. Phys. \textbf{76}, 785 (2004).
\bibitem{RC} D.R. Reichman and P. Charbonneau, J. Stat. Mech. P05013 (2005).
\bibitem{Szamel2003} G. Szamel, Phys. Rev. Lett. \textbf{90}, 228301 (2003).
\bibitem{WuCao} J. Wu and J. Cao, Phys. Rev. Lett. \textbf{95}, 078301 (2005).
\bibitem{Janssen2015} L.M.C. Janssen and D.R. Reichman, Phys. Rev. Lett. 
\textbf{115}, 205701 (2015).
\bibitem{Note1} See Refs. \cite{Franz2005,Dzero2005,Franz2007}
for calculations of activation barriers, Ref. \cite{Schweizer2005} for 
a single-particle-based theory of activated hopping, later generalized to
include ``elastically collective'' effects in Ref. \cite{Schweizer2014}
and Refs. \cite{Rizzo2014,Rizzo2016} for a dynamical field theory 
that predicts that the mode-coupling transition is avoided.
\bibitem{Franz2005} S. Franz, J. Stat. Mech. P04001 (2005).
\bibitem{Dzero2005} 
M. Dzero, J. Schmalian, and P. G. Wolynes, Phys. Rev. B \textbf{72},
100201(R) (2005).
\bibitem{Franz2007} S. Franz, J. Stat. Phys. \textbf{2126}, 765 (2007).
\bibitem{Schweizer2005} K.S. Schweizer, J. Chem. Phys. \textbf{123}, 244501 (2005).
\bibitem{Schweizer2014} S. Mirigian and K.S. Schweizer, 
J. Chem. Phys. \textbf{140}, 194506 (2014).
\bibitem{Rizzo2014} T. Rizzo, EPL \textbf{106} 56003 (2014).
\bibitem{Rizzo2016} T. Rizzo, Phys. Rev. B \textbf{94}, 014202 (2016).
\bibitem{MP1}
M. M\'ezard and G. Parisi, J. Phys. A: Math. Gen. \textbf{29} 6515 (1996).
\bibitem{MP2}
M. M\'ezard and G. Parisi, Phys. Rev. Lett. \textbf{82} (1999) 747; J. Chem. Phys.
\textbf{111} 1076 (1999).
\bibitem{PZ1} G. Parisi and F. Zamponi, J. Chem. Phys. \textbf{123}, 144501 (2005).
\bibitem{PZ2} G. Parisi and F. Zamponi, Rev. Mod. Phys. \textbf{82}, 789 (2010).
\bibitem{JacquinZamponi} H. Jacquin and F. Zamponi, 
J. Chem. Phys. \textbf{138}, 12A542 (2013).
\bibitem{Note2} It is possible to formulate static approach
that is consistent with the mode-coupling theory if instead of the free-energy-like
functional one starts from a replicated Ornstein-Zernicke equation and uses a 
specific closure approximation for the replica off-diagonal direct correlation
function, see Ref. \cite{mctstatic}. The closure consistent with the 
mode-coupling theory cannot be obtained from a free-energy-like functional,
as shown explicitly in Ref. \cite{JacquinZamponi}.
\bibitem{mctstatic} G. Szamel, Europhys. Lett. \textbf{91}, 56004 (2010).
\bibitem{RFOTorg}
T. R. Kirkpatrick, D. Thirumalai, and P. G. Wolynes, Phys. Rev.
A \textbf{40}, 1045 (1989).
\bibitem{RFOTrec} V. Lubchenko and P. G. Wolynes, Annu. Rev. Phys. Chem.
\textbf{58}, 235 (2007); \textit{Structural Glasses and Supercooled Liquids: Theory,
Experiment, and Applications}, edited by P. Wolynes and
V. Lubchenko (Wiley, New York, 2012).
\bibitem{PUZbook} G. Parisi, P. Urbani, and F. Zamponi, 
\textit{Theory of Simple Glasses: Exact Solutions in Infinite Dimensions} 
(Cambridge University, Cambridge, England, 2020).
\bibitem{highdstatic} J. Kurchan, G. Parisi, F. Zamponi J. Stat. Mech. P10012 (2012);
J. Kurchan, G. Parisi, P. Urbani, F. Zamponi,
J. Phys. Chem. B \textbf{117}, 12979 (2013); 
P. Charbonneau, J. Kurchan, G. Parisi, P. Urbani, F. Zamponi,
J. Stat. Mech. P10009 (2014).
\bibitem{highddynamic} T. Maimbourg, J. Kurchan and F. Zamponi,
Phys. Rev. Lett. \textbf{116}, 015902 (2016); 
E. Agoritsas, T. Maimbourg, and F. Zamponi, J. Phys. A: Math. Theor. \textbf{52},
144002 (2019).
C. Liu, G. Biroli, D. Reichman, and G. Szamel,
Phys. Rev. E \textbf{104}, 054606 (2021).
\bibitem{SchmidSchillingPRE} B. Schmid and R. Schilling, 
Phys. Rev. E \textbf{81}, 041502 (2010).
\bibitem{IkedaMiyazakiPRL2010} A. Ikeda and K. Miyazaki,
Phys. Rev. Lett. \textbf{104}, 255704 (2010).
\bibitem{Manacorda} A. Manacorda, G. Schehr, and F. Zamponi, 
J. Chem. Phys. \textbf{152}, 164506 (2020).
\bibitem{FP1} S. Franz and G. Parisi, J. Phys. I (France) \textbf{5}, 1401 (1995).
\bibitem{FP2} S. Franz and G. Parisi, Phys. Rev. Lett. \textbf{79}, 2486 (1997).
\bibitem{DDFTreview} M. te Vrugt, H. L\"owen and R. Wittkowski,
Adv. Phys. \textbf{69}, 121 (2020).
\bibitem{Note3} Systems with microscopic Newtonian and Brownian dynamics
exhibit the same glassy dynamics features
\cite{SzamelFlenner2004}. Assuming Brownian dynamics eliminates
an additional approximation needed to eliminate the velocities of the particles.  
\bibitem{SzamelFlenner2004} G. Szamel and E. Flenner, 
Europhys. Lett. {\bf 67}, 779 (2004).
\bibitem{SL} G. Szamel and H. L\"{o}wen, Phys. Rev. A \textbf{44}, 8215 (1991).
\bibitem{ResiboisLebowitz} 
P. R\'esibois and J.L. Lebowitz, J. Stat. Phys. \textbf{12}, 483 (1975).
\bibitem{BlawzdzCichocki} J. B\l awzdziewicz and B. Cichocki, 
Physica A \textbf{127}, 38 (1984).
\bibitem{SzamelLeegwater} See the discussion in the Appendix of
G. Szamel and J.A. Leegwater, Phys. Rev. A \textbf{46}, 5012 (1992).
\bibitem{Kawasaki} K. Kawasaki, Ann. Phys. \textbf{61}, 1-56 (1970) and 
references therein.
\bibitem{MarconiTarazona} 
U. M. B. Marconi and P. Tarazona, J. Chem. Phys. \textbf{110}, 8032 (1999).
\bibitem{ArcherEvans} A. J. Archer and R. Evans, J. Chem. Phys. \textbf{121}, 4246 
(2004).
\bibitem{EspanolLoewen} P. Espa\~nol and H. L\"owen, J. Chem. Phys. \textbf{131}, 
244101 (2009).
\bibitem{DuftyRodriguez} J.W. Dufty and R.F Rodriguez, 
J. Stat. Phys. \textbf{33}, 261 (1983).
\bibitem{Hauge} E. Hauge, Phys. Rev. Lett. \textbf{28}, 1501 (1972).
\bibitem{FranzQE1} S. Franz, G. Parisi and P. Urbani, 
J. Phys. A: Math. Theor. \textbf{48} 19FT01 (2015).
\bibitem{FranzQE2} S. Franz, G. Parisi, F. Ricci-Tersenghi and P. Urbani, 
J. Stat. Mech. P10010 (2015).
\bibitem{Note5} The dynamic density functional theory 
evolution equation is usually written in the direct 
space\cite{DDFTreview,MarconiTarazona,ArcherEvans,EspanolLoewen}. We wrote it
in the Fourier space to make the similarity with our Eq. \eqref{dcfeom4} explicit.
\bibitem{Cammarota2010} C. Cammarota, A. Cavagna, I. Giardina, G. Gradenigo, 
T.S. Grigera, G. Parisi, and P. Verrocchio, 
Phys. Rev. Lett. \textbf{105}, 055703 (2010).
\bibitem{KawasakiGunton} K. Kawasaki and J.D. Gunton, 
Phys. Rev. A \textbf{8}, 2048 (1973).
\bibitem{SchweizerSaltzman} K. S. Schweizer and E. J. Saltzman, 
J. Chem. Phys. \textbf{119}, 1181 (2003).
\end{thebibliography}

\end{document}